# Influence of Thermal Cycling on Giant Magnetocaloric Effect of Gd$_5$Si$_{1.3}$Ge$_{2.7}$ Thin Film

A L Pires[1,2], J H Belo[1], I T Gomes[1], R L Hadimani[3,4], D L Schlagel[4], T A Lograsso[4,5], D C Jiles[3,4], A M L Lopes[1,2], J P Araújo[1*], A M Pereira[1*]

[1]*IFIMUP and IN - Institute of Nanoscience and Nanotechnology, Departamento de Física e Astronomia da Faculdade de Ciências da Universidade do Porto, Rua do Campo Alegre, 687, 4169-007 Porto, Portugal.*

[2] *CFNUL - Centro de Física Nuclear da Universidade de Lisboa, Av. Prof. Gama Pinto, 2, 1649-003 Lisboa, Portugal.*

[3]*Department of Electrical and Computer Engineering, Iowa State University, Ames, Iowa 50011, USA*

[4]*Ames Laboratory, US Department of Energy, Iowa State University, Ames, Iowa 50011, USA*

[5]*Division of Materials Science and Engineering, Ames Laboratory, Ames, Iowa 50011, USA*

In the lifetime of a magnetic refrigerator the materials are subjected to millions of thermal magnetic cycles, so it is of the utmost importance to predict the behavior of such materials when subjected to related work conditions. Thus, in this work we investigated the influence of thermal cycling in the microstructure, magnetic phase transition and magnetic entropy change of a Gd$_5$Si$_{1.3}$Ge$_{2.7}$ thin film up to 1000 cycles. In a first stage, till 450 cycles, the hysteresis area of the magnetization curves as a function of temperature was found to decrease 16% with thermal cycling, due to an arresting of the O(II) phase caused by internal strain. Nonetheless, after 1000 cycles there is a clear loss of the magneto-structural transition. For all thermal cycles the O(I) phase remains unchanged. Therefore, prolonged thermal cycling leads to a strong reduction of the O(II) phase, attributed to an internal pressure build-up, caused by the large number of expansion/compression cycles that the O(II) phase undergoes across the magnetostructural transition.

In the last decade there has been an increasing demand towards clean energy and energy efficient devices, which will replace the more polluting and less efficient ones, such as refrigeration systems running on gases that contribute to the greenhouse effect [1]. In this framework, magnetic refrigeration systems emerge as the most prominent candidates, since they are more efficient and are not considered harmful to the environment [2,3]. The magnetic refrigeration at room temperature is based on the magnetocaloric effect (MCE), which occurs in magnetic materials that exhibit a large magnetization change under the application of a magnetic field. They transform from a magnetically disordered state to an ordered state (at a temperature close to its critical temperature), and consequently large magnetic entropy changes occur. In recent years, techniques such as doping and thermal treatments have been used to enhance the magnetocaloric effect [4–10]. It is consensual that the magnetic materials undergoing first-order phase transitions are the ones exhibiting larger MCE, which is a key-parameter for their application in refrigeration systems. The $Gd_5(Si_xGe_{1-x})_4$ [11–13], $La(Fe_xSi_{1-x})_{13}$ [14,15], $MnP_{1-x}As_x$, compounds belong to the group of first-order magnetic materials that exhibit a giant magnetocaloric effect (GMCE) [4,6,16]. However one drawback of these compounds is that their physical properties change when subjected to magnetic/thermal cycling [15,17]. One of these changes is related to their MCE and in particular its magnitude during the cycling process, which is crucial for the magnetic refrigerator's performance. In magnetic refrigeration many parameters can influence the performance of these systems over time. For instance, for $LaFe_{11.6}Si_{1.4}$ it has been shown that an increase of the magnetic field promotes a reduction of thermal hysteresis upon increasing temperature [15]. The authors associated this result to the presence of a first-order magnetic transition that includes metastable states in addition to thermal equilibrium states. Also, properties such as electrical resistance [18–21], thermopower [22] and specific heat [23] have been reported as dependent of thermal cycling on $Gd_5(Si_xGe_{1-x})_4$ alloys. One of the ways to recover from these properties changes is by low temperature annealing, as suggested by Hadimani et al [24]. Phenomenon as disordered atomic structure and aging also have been associated to the reduction of the both transition temperature and magnetic entropy change [25]. However, all studies performed on these compounds have been focused in the bulk form or in micrometric scale samples. Nowadays, driven by the prospect of producing miniaturized devices, from micro-refrigerators to magnetic/pressure/temperature sensors [26–29], the interest on the study of MCE in materials with reduced dimensions has been growing considerably. Additionally, MC thin films will be readily adapted to integrate micro-magnetocaloric processes inside micro-electronic circuits [30], because of their advantageous higher surface to volume/ratio. Ball milling technique allows to obtain nanometer size particles which have higher surface/volume ratio leading to a decrease of the overall magnetization values. This phenomenon can be due both to a magnetic moment blocking mechanism and an amorphization of the particles structure [31,32]. Recently, Zhang et al [33] showed that both temperature and applied magnetic field cycles change the hysteresis of the MnAs nanocrystals. Therefore, it is important to understand the behavior of magnetocaloric materials at the nanoscale when they undergo a large number of thermal cycles, simulating its operation in a refrigeration system [23]. The aim of this work is to report the effect of consecutive thermal cycles on the microstructure, magnetization and magnetic entropy changes of a $Gd_5Si_{1.3}Ge_{2.7}$ thin film.

* Corresponding authors:
Dr. André Pereira, Tel.: +351 220402369; fax: +351220402406. *E-mail address:* ampereira@fc.up.pt; and
Dr. João Pedro Araújo, Tel.: +351-22 04 02 362. *E-mail address:* jearaujo@fc.up.pt.

The preparation details of the $Gd_5Si_{1.3}Ge_{2.7}$ thin film on a $SiO_2$-covered Si substrate using femtosecond pulsed laser ablation were described in Ref. [27]. The thermal cycling was performed by immersing the $Gd_5Si_{1.3}Ge_{2.7}$ thin film (area=8.2 mm$^2$, thickness 763 ± 25 nm) in a liquid nitrogen bath. In each cycle, the film was immersed for 60 seconds in order to ensure thermal equilibrium with the bath and afterwards it was removed immediately. This cycle was repeated 1000 times and between consecutive dips the sample rested at room temperature during 60 seconds. Scanning Electron Microscopy (SEM) was used to evaluate the microstructure of the thin film before starting cycling and after 50 and 450 thermal cycles. X-ray diffraction spectra and magnetic measurements were also performed before cycling and after 50, 200, 250, 450 and 1000 cycles in a: MiniFlex600 Rigaku and commercial (Quantum Design MPMS-5S) Superconducting Quantum Interference Device (SQUID) magnetometer, respectively. The magnetic measurements performed were isothermal M(H) curves up to 50 kOe. The magnetic entropy changes [-$\Delta S_m(T)$] were estimated through the application of the Maxwell relation [34]. Temperature dependence of magnetization was measured before and after 450 and 1000 cycles.

Figure 1 shows the SEM of the thin film top view, obtained for the same specific region before cycling and after 50 and 450 cycles. From Figures 1 a1) - b1), it is observed that the micro-morphology of the thin film presents a clustering of spherical particles that remains almost unaltered during the thermal cycling for the sample after 50 cycles. For the sample after 450 cycles (Figure 1 c1) the morphology remains the same but is not possible to identify the same initial spot, i.e. the thermal cycling changes the surface of the thin films. We attribute such small changes to the induced strain promoted by the thermal cycles that through the nanoparticles expansion/shrinking can promote their slight movement along the surface. This is patent when performing the size histogram and using a Log-Normal distribution, where similar mean sizes of 108 nm before cycling, and 112 nm after 50 and 450 cycles are obtained, Figure 1a2) – c2), within the error. The major change is on the full width at half maximum of the distribution that decreases with the cycling, showing a narrower size distribution.

The thin film magnetization behavior as a function of temperature, before and after thermal cycles, is shown in Figure 2a). It can be observed that the $Gd_5Si_{1.3}Ge_{2.7}$ thin film exhibits two magnetic transitions [9,27,35]. At T~249 K a magnetic transition occurs, corresponding to a pure paramagnetic to ferromagnetic second-order transition (orthorhombic O(I) crystal structure is retained). At lower temperature, T~197 K, a structural and a magnetic transition occur simultaneously - magnetostructural transition (MST) - from a paramagnetic orthorhombic O(II) to a ferromagnetic orthorhombic O(I) phase. Furthermore, in this first-order transition, thermal hysteresis is observed, where the transition temperatures for heating and cooling are ~197 K and ~187 K, respectively [6,35].

After 450 cycles, the magnetization curves as a function of temperature, M(T), exhibit a similar behavior as the as-deposited thin film, (Figure 2a)) overlapping in the [210-300] K temperature range. The thermal cycling does not seem to affect the second-order transition that occurs at 249 K. The major difference arises at lower temperatures (~197 K on heating) because thermal cycling causes a decrease in the amount of phase that undergoes the MST (~ 16% hysteresis area reduction). This statement is confirmed by the M(H) loops (not shown here) obtained at same temperature leading to a 21% decrease of the magnetic hysteresis. Such decrease is an indication that thermal cycling is mainly actuating on the MST. When increasing the number of thermal cycles to 1000 cycles, the second-order phase transition still occurs at 249 K. However, it is possible to observe the complete disappearance of the MST at lower temperature.

In order to study the atomic structure, XRD analysis was carried out at room temperature (Figure 2b). The XRD patterns show peaks corresponding to the O(I) phase ($Gd_5Si_4$) [36] and O(II) phase ($Gd_5Ge_4$) [37] identified with "*" and "+", respectively. Upon comparison, it is observed that there is an overall peak-intensity decrease after cycling. However such decrease is more drastic for the O(II) phase peaks, such as (3 1 1) and (1 6 4), as their intensities are clearly reduced after 450 cycles and disappear completely after 1000 cycles. This corroborates the hypothesis of O(II) phase loss, as suggested by the magnetization results. At lower angles a shift of the peaks is also observed. This shift can be a signal of the induced strain after cycling. In particular, the XRD pattern for the sample after 1000 cycles reveals the absence of the O(II) phase peaks. Only one peak of the O(I) phase remains present in the XRD (reflection (1 4 3)). And other peak appears after 450 cycles and increase after 1000 cycles (reflection (2 1 1)).

The saturation moments ($\mu_S$) were calculated from the magnetization curves at 5 K considering the $Gd_5Si_{1.3}Ge_{2.7}$ bulk density (7.7 gcm$^{-3}$) [6]. The estimated values for the sample before and after 450 and 1000 cycles are 6.7 ± 0.5, 6.2 ± 0.5 $\mu_B$ and 6.0 ± 0.5 respectively. These values are close to the theoretical 7 $\mu_B$ saturation magnetic moment of pure $Gd^{3+}$ ions [6].

The magnetic entropy variation was estimated by integrating the Maxwell relation on the M(H) curves for before and after 50, 200, 250, 450 and 1000 cycles, as is shown in Figure 3a) in the [170, 225] K temperature range. For all curves the peak value of the magnetic entropy change, $-\Delta S_m^{max}$, occurs around the same temperature, T ~192.5 K, as a consequence of the MST occurring at this constant temperature [27]. Concerning the dependence of $\Delta S_m$ magnitude as a function of the number of cycles, before cycling, the thin film has a maximum magnetic entropy change $-\Delta S_m^{max}$ ~ 64 mJK$^{-1}$cm$^{-3}$, then decreases after 50, 200 and 450 cycles by 10% (58 mJK$^{-1}$cm$^{-3}$), 15% (55 mJK$^{-1}$cm$^{-3}$) and 17% (53 mJK$^{-1}$cm$^{-3}$), respectively. At the end of the 450 cycles, a 19% reduction of the $-\Delta S_m^{max}$ value was observed (~52 mJK$^{-1}$cm$^{-1}$). However, after 1000 cycles the magnetic entropy change reduces to ~12 mJK$^{-1}$cm$^{-1}$ (corresponding to a reduction of 81%). After 1000 cycles, the O(II) phase, responsible for the magnetostructural transition, seems to undergo an internal pressure build-up that causes a drastic decrease on the $\Delta S_m$ magnitude.

Concerning the refrigerant capacity, $RCP_{FWHM}$, we observed a decreased with the number of cycles, with a maximum variation for 450 cycles ($RCP_{FWHM\ Before\ Cycling}$=1239 mJcm$^{-3}$; $RCP_{FWHM\ After\ 450\ Cycling}$=1127 mJcm$^{-3}$). After 1000 cycles the estimation of $RCP_{FWHM}$ was no longer possible.

Analysing the obtained results, namely magnetic, structural and morphological, it is clear that the thermal cycling is playing an important role on driving the structural transition. This is evident on the magnetization curves, M(T) - (Figure 2a), due to the systematic decrease in the hysteresis area, until the complete disappearance after 1000 cycles. In these materials, the structural transition induces an abrupt volume change (1.2%), where the smallest volume phase stabilizes at lower temperatures, i.e., [V(O(I)) < V(O(II))] [26]. During the cooling process the appearance of free space due to the contraction of the unit cells is expected, thus releasing internal strain that can be formed during the thin film growth. For the reverse process, i.e. during heating, a significant increase in the unit cell volume will lead to a stress increase arising from the volume cell expansion, thus originating the arrest of the structural transition. In fact, this mechanism has been referenced several times for the bulk materials to explain changes in their properties, namely morphological, magnetic and transport [21]. Also the same explanation was used for other family of systems showing drastic volume changes [27]. In the $R_5(Si,Ge)_4$ family bulk compounds, we highlight the works from Casanova and co-authors and Perez and co-authors which demonstrate that the metastability, responsible for the thermal hysteresis, is reduced when the number of thermal cycles increases [23,38]. Furthermore, a previous study on the $Gd_5(Si_{0.1}Ge_{0.9})_4$ electrical resistivity behavior has also shown a thermal hysteresis area reduction as the number of cycles increase: the number of defects, the spin disorder, domain reorientation or the preferential Si/Ge sites were assigned as the major causes leading to the metastability reduction [20,21]. The local behavior of the magnetic relaxation (across the field increase and decrease) of the $Gd_5Ge_4$ first-order transition has also been investigated [39]. It was concluded that the transition at lower temperatures is arrested and a metastable equilibrium (disorder-influenced first-order transition) is obtained, leading to a significant decrease in magnetic relaxation. On the other hand, at high temperatures, only a slight decrease was observed due to a faster process towards the metastable equilibrium. At these temperatures the activation over the excitation strain energy barriers is facilitated [39–41]. In fact, the system attains the equilibrium in a faster and more homogeneous way, i.e., tends to reduce the dissipated energy and therefore to reduce the hysteresis [38]. On other systems exhibiting a MST, such as in $Ni_{54.3}Mn_{20.1}Ga_{25.6}$ alloys [17], a decrease of the magnetic moment with the number of cycles was also detected and associated with an arrest of the small volume phase (the austenite phase in that case), caused by local strain. This phenomenon is also extended to $La(Fe,Si)_{13}$ family of compounds where a decrease of the MCE in a few cycles was observed in the bulk form [29]. In other study, the stress relaxation, accompanied by a short-range order structural change, is responsible for the decrease of the $\Delta S_m$ [25].

Therefore, we believe that the stress and strain generated during the cycling, across first-order phase transition is responsible for the decrease of the entropy as a function of thermal cycles. The decrease of the particle size can also be responsible for these reduction [31]. During the first 450 thermal cycles this stress and strain seem to promote an arrestment of the O(II) which disables its complete transformation into O(I) phase, consequently reducing the overall magnetic entropy change. After 450 cycles, the O(II) phase, previously arrested, is drastically reduced due to the intense build up stress that is subjected to. So, from 450 cycles to 1000 cycles we suggest that a phenomenon such as structural disorder promotes the loss of the O(II) phase (confirmed by magnetic and structural data) and consequently the absence of the magnetostructural transition. This loss is accompanied with an increase of paramagnetism in the samples as can be observed in Inset of Figure 3a). In fact, the M(H) curve at 5K after 1000 cycles shows a distinct behavior corresponding to a clear decrease of ferromagnetism. The XRD data corroborate this assumption as all diffraction peaks disappear after 1000 cycles. Recently Zhang et al [33], performed cooling cycles on MnAs nanocrystals (down to liquid nitrogen), and have also reported a loss of hysteresis , loss of MST, promoted by temperature and field cycles. It is important to remark that, in comparison with the bulk form, where the crystallographic grain surface area is several micrometer-square, the nanoparticles surface can be up to three orders of magnitude lower, which enhances (by three orders of magnitude) the stress that these nanoparticles are subjected to along the magnetostructural transition. This behavior has a parallel with the ball milling process used for a top-down approach to produce $Gd_5(Si,Ge)_4$ nanoparticles where disorder and amorphization are also associated with a decrease of magnetic ordering and consequent decrease of magnetocaloric effect. [25,31,42]. Also, Vishnoi et al [42], reported the degradation of the first-order transition and the amorphization on the Ni50Mn35.6Sn14.4 films with the increase of ion irradiation, i.e, increasing stress.

In summary, with this work we report the effects of thermal cycling on a $Gd_5Si_{1.3}Ge_{2.7}$ thin film properties. The magnetization and structural investigation in this thin film showed a decrease/disappearance of the O(II) phase. After 450 cycles, it is observed that the phase is arrested due to the strong strain that is created after consecutive MST. After 1000 cycles a complete loss of the O(II) phase is observed, attributed to disorder and amorphization of the sample. In the future, studies using acoustic emission may be useful to better understand these results, in particular for the investigation of the kinetics of the structural transition and the crystallographic phase evolution in such thin films. Also, the behavior of freestanding nanoparticles (not constrained in a substrate) needs to be studied. Finally, a comparison with a non-granular thin film would give also great insight for the full comprehension of the operating stress/strain mechanisms.

**ACKNOWLEDGMENTS**

The authors acknowledge FCT for financial support through the projects: PTDC/CTM-NAN/115125/2009, EXPL/EMS-ENE/2315/2013 and FEDER/POCTIn0155/94. J H Belo thanks FCT for the Grant SFRH/BD/88440/2012. A M Pereira, I T Gomes and A M L Lopes acknowledge the project NORTE-070124-FEDER-000070 for the financial support. A L Pires thanks for the Grant: PEst-OE/FIS/UI0275/2014 and Incentivo/FIS/UI0275/2014. Work at Ames Laboratory was supported by the U.S. Department of Energy, Office of Basic Energy Sciences, Division of Materials Science and Engineering. Ames Laboratory is operated for DOE by Iowa State University under Contract No. DE-AC02-07CH11358.

# Figure 1

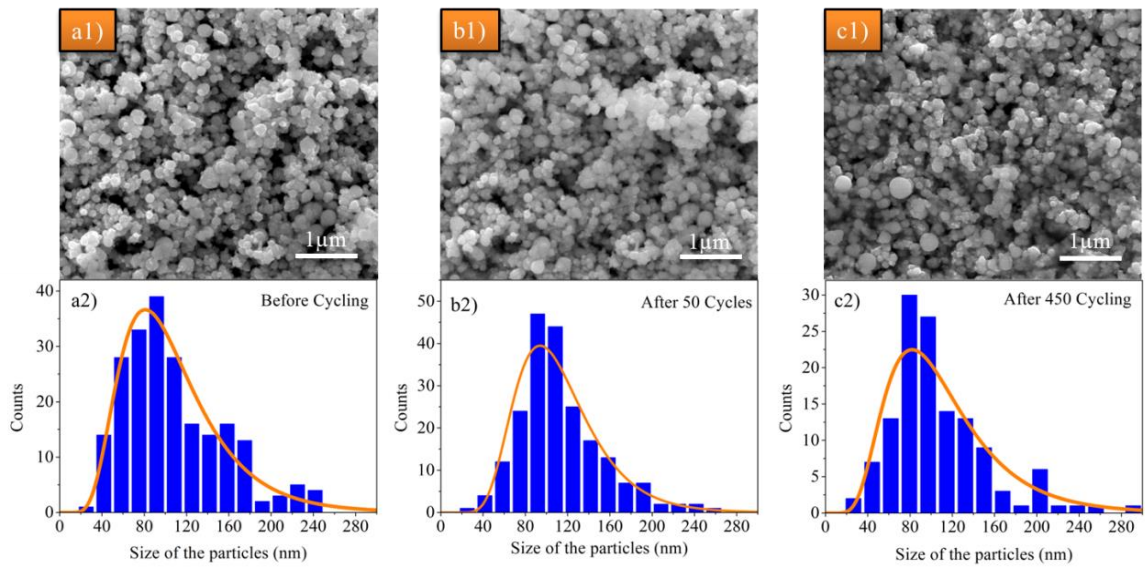

Figure 1: SEM: (a1) before thermal cycling; (b1) after 50 thermal cycles and (c1) after 450 thermal cycles; and the corresponding histograms (a2, b2 and c2) with Log-Normal distribution.

# Figure 2

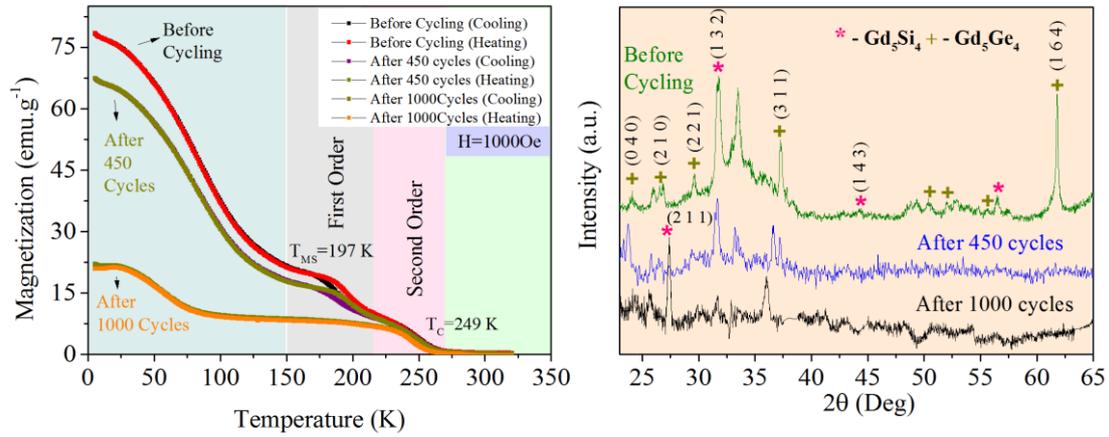

Figure 1 – (color online) a) Magnetization as a function of temperature for the $Gd_5Si_{1.3}Ge_{2.7}$ thin film: before cycling, and after 450 and 1000 thermal cycles in cooling and heating with a field of 1000 Oe; b) X-ray diffractograms of the samples: before cycling (green), after 450 cycles (blue) and after 1000 cycles (black). The peaks marked with [*] correspond to the O(I) $Gd_5Si_4$ -like structure and the ones marked with [+] correspond to O(II) $Gd_5Ge_4$ -like structure.

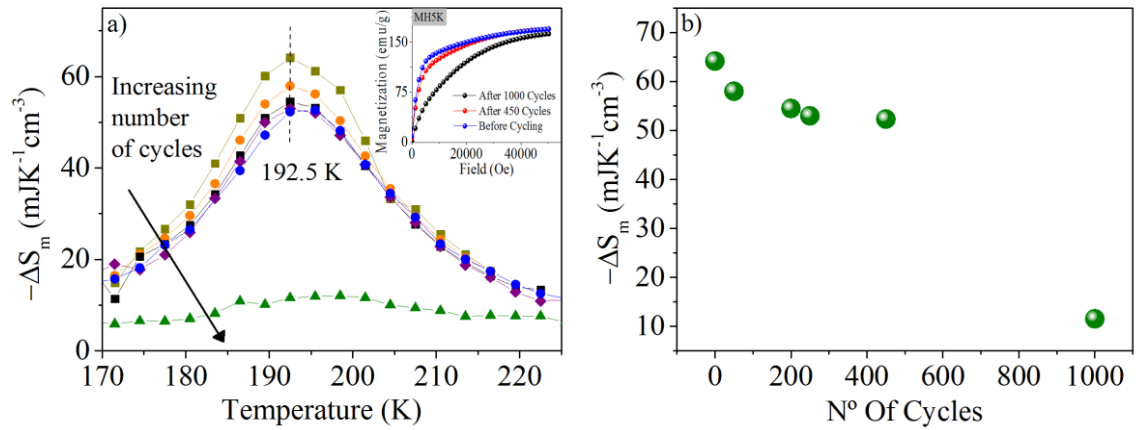

Figure 2 – (color online) a) Magnetic entropy change in the [171-225] K temperature range of the $Gd_5Si_{1.3}Ge_{2.7}$ thin film before cycling and after 50, 200, 250, 450, and 1000 cycles under an applied magnetic field change of ΔH=50 kOe Inset: Magnetization versus Field for samples before cycling and after 450 and 1000 cycles at 5 K; b) Representation of the magnetic entropy change peak value as a function of the number of cycles.